# An Agentic ENF Framework for Forensic Video Authentication Against Generative AI

Ryan Wyton and Bacha Rehman

***Abstract*: Generative models now synthesise video that is hard to distinguish from genuine footage, and learned detectors that key on a generator's artefacts degrade sharply on unseen models. This paper develops a forensic framework that authenticates video by a signal that generation does not control: the electric network frequency (ENF), the mains-driven hum in audio and flicker in video. A unified two-channel detector extracts the ENF from audio and from rolling-shutter video through one second-harmonic estimator and a shared network; a synthetic-injection training method, driven by an Ornstein-Uhlenbeck drift model and anchored to real recordings, overcomes the scarcity of labelled ENF data; a reference-matching channel verifies a claimed time of recording and recovers an unknown one; and a two-axis agentic orchestrator fuses recording authenticity with content-manipulation evidence under an asymmetric rule, escalating expensive detectors only when needed. A central finding reframes the task: an ENF-band flicker is present in generated video at a prominence indistinguishable from real footage, so its presence is not proof of authenticity, and authentication must rest on consistency with an independent reference rather than on presence. On held-out real recordings the detector accepts genuine audio and video at 93% and 100% against naive baselines of 53% and 15%; claimed-time authentication reaches an area under the curve of 1.0; and the agentic cascade matches a flat ensemble while cutting wall-clock time by 54% and frames decoded by 64%, without loss of accuracy.**

## I. Introduction

Generative models now synthesise video that is difficult to distinguish from genuine footage [1], and its misuse for disinformation and fraud has made detection a pressing forensic problem [2]. Learned detectors, however, key on the statistical fingerprints of the generators they are trained on, a failure mode characterised for generative adversarial network (GAN) images [3]. It carries over to video: on a recent six-million-clip benchmark the strongest detectors reach only around 80% accuracy overall, and on generators held out of training they fall to near or below chance [4]. Detection framed as the classification of synthesis artefacts is therefore an arms race the detector is losing. A more durable basis for authentication is a signal that generation does not control. The electric network frequency (ENF), the small drift of the mains supply about its nominal 50 or 60 Hz value, is such a signal: it is shared across a power grid at any instant, it fluctuates in a way characteristic of a region and time, and it is imprinted on recordings as a hum in audio and, through rolling-shutter capture, as a flicker in video at twice the grid frequency [5]-[7]. A genuine recording carries a coherent, naturally drifting ENF that can be checked against an independent record of the grid; a synthesised recording does not measure a live grid.

The premise from which this work began, that synthetic video would simply lack an ENF and could be flagged by its absence, did not survive contact with the evidence. Modern text and image-to-video generators reproduce optical flicker at a prominence comparable to real footage, and its apparent grid is set by the generator's characteristic frame rate rather than measured from a live supply [8]. Presence of an ENF-band line is therefore not, by itself, evidence of a genuine recording, and a single short clip cannot recover even the nominal grid, which is confounded by the camera frame rate. The problem is accordingly reframed: from detecting a fake by the absence of a signal, to authenticating a genuine recording by the consistency of its ENF with an independent reference.

This paper develops a forensic framework that operationalises a shift from detecting fakes by the absence of a signal to authenticating recordings by the consistency of one. Its contributions are as follows. First, a unified two-channel detector extracts the ENF from both audio and rolling-shutter video through a single second-harmonic estimator and a shared network, so one model serves both modalities. Second, a synthetic-injection training methodology, driven by an Ornstein-Uhlenbeck model of grid drift and anchored to a small set of real recordings, overcomes the scarcity of labelled ENF data and defines a six-class taxonomy: a genuine class and five modes of ENF inconsistency. Third, a reference-matched authentication channel verifies a claimed time of recording and recovers an unknown one, and, where no external grid reference is available, can cross-check a clip's optical and audio ENF against each other, turning ENF from a presence cue into a timestamp and a consistency test. Fourth, a two-axis agentic orchestrator fuses recording-authenticity evidence from ENF with content-manipulation evidence from existing visual baselines under an asymmetric rule: a detected cue is evidence of authenticity, its absence is not evidence of manipulation, and a confident content-fake overrides a genuine scene ENF. Expensive detectors are escalated only when the cheap tier is inconclusive, and the decision threshold is calibrated to minimise false accusation rather than to maximise accuracy.

The remainder of the paper is organised as follows. Section II reviews related work. Section III sets out the ENF

Manuscript submitted 15 October 2026. This work received no external funding and was carried out as part of the first author's MSc dissertation at Southampton Solent University. R. Wyton and B. Rehman are with Southampton Solent University, Southampton, U.K. (e-mail: 0wytor22@solent.ac.uk; bacha.rehman@solent.ac.uk). Code and data is available at https://github.com/RyanSolent95/enf-video-forensics.

background and the threat model. Section IV describes the framework. Section V details the experimental setup, Section VI reports results, and Section VII discusses limitations before Section VIII concludes.

## II. Related Work

### A. ENF forensics

ENF has been used in multimedia forensics for two decades, initially in audio and later in video. Early work established that mains hum recorded alongside speech encodes a temporal signature that can be matched against a grid database [5]. Subsequent studies demonstrated that this same signal, which is imprinted optically through rolling-shutter capture under flickering illumination, can be extracted directly from digital video [6], [7]. Extraction from rolling-shutter footage has since been refined, for example by periodic zero-padding that corrects the readout geometry [9], by detailed analysis of the rolling-shutter effect on the recovered signal [10], and by noise-resilient multi-harmonic estimators [11]. Beyond timestamping, the ENF supports region-of-recording identification [12], detection of the presence of an ENF trace in a clip [13], camera forensics from power signatures [14], and time-of-recording verification by spectrum combining [15]; the area is surveyed in [16]. These methods largely assume a genuine recording and ask when or where it was made, whereas the present work asks whether a recording is genuine at all, in a setting where the adversary is a generator.

### B. ENF for tampering and deepfake detection

A smaller body of work uses the ENF to expose manipulation: inconsistencies in the trace can reveal splicing, learned models combine ENF features with sequence networks for video tamper detection [17], and decentralised schemes use the ENF as an authentication token against deepfakes [18]. These approaches treat the presence or continuity of an ENF as the mark of authenticity. This work departs from them in two ways: it shows that presence alone no longer suffices, because generated video shows an ENF-band flicker of comparable prominence, and it bases authentication on consistency with an independent reference rather than on the trace's mere presence.

### C. AI-generated video detection and its fragility

The dominant approach to detecting generated video is a classifier trained on the artefacts of known generators [19], using spatial, temporal or spatio-temporal features and, more recently, vision-language models. Such detectors are accurate in distribution but degrade sharply on generators unseen in training, a collapse analysed for GAN images [3] and confirmed for video, where on a recent benchmark [4] the strongest video-level detectors reach only around 80% accuracy and fall to near or below chance on out-of-distribution generators. This fragility under domain shift motivates grounding authentication in a physical signal that generation does not control, rather than in the shifting statistics of the generators themselves.

### D. Physical and physiological cues

A complementary line of work authenticates video through physical or physiological signals that a manipulation must also reproduce. Remote photoplethysmography (rPPG) recovers a pulse from subtle skin-colour changes, whose disruption can indicate a synthesised or swapped face [20]; lip-sync consistency checks the correspondence between speech and mouth motion [21]; and broader reviews of forensic video analysis survey these and related cues alongside provenance and metadata [22]. The framework here uses these as its content axis, but treats them, like the ENF, asymmetrically: a detected cue is evidence of authenticity, whereas its absence, common on compressed or low-quality footage, is not evidence of manipulation. The contribution is not any single detector but the two-axis fusion of a physical recording-authenticity signal with these content cues under an agentic policy.

## III. Background and Threat Model

### A. The electric network frequency

Alternating-current mains are nominally 50 Hz across much of the world and 60 Hz in the Americas, but their instantaneous frequency drifts by small amounts as supply and demand vary [5]. As a synchronous grid is tightly coupled, this drift is shared, to within milliseconds, across the whole interconnection at any instant, which is what makes it a forensic reference: a recording's ENF can be matched against an independently logged grid trace to place it in time, and in principle by region. However, for a single short clip the regional read is unreliable, being confounded by the camera frame rate [8]. Mains-powered lighting modulates its output at twice the supply frequency (1), so a scene lit this way fluctuates at 100 or 120 Hz. This reaches a recording in two ways: as a hum in the audio, and, when a rolling-shutter sensor captures the scene, as a banding flicker in the video.

$$f_{\text{flicker}} = 2\, f_{\text{grid}} \tag{1}$$

### B. Ornstein-Uhlenbeck model of grid drift

The drift is not a random walk but is mean-reverting: it wanders around the nominal value and is pulled back by grid regulation. The drift is therefore modelled as an Ornstein-Uhlenbeck (OU) process, a continuous-time process with a reversion rate, a long-run mean and a volatility, defined by the stochastic differential equation (2), where $\theta$ is the reversion rate, $\mu$ the long-run mean, $\sigma$ the volatility, and $W_t$ a Wiener process. For simulation the state is advanced with the exact discrete-time transition (3), a step of $\Delta$ with Gaussian innovation $\varepsilon_t$ whose variance keeps the process stationary. Parameters are estimated from real mains measurements by fitting the equivalent first-order autoregressive form (4): regressing the mean-removed trace $Y_t$ on its lag gives $\hat{b}$, from which the reversion rate $\hat{\theta} = -\ln \hat{b} / \Delta$ and the stationary standard deviation $\sigma_\infty = \sigma/\sqrt{2\theta}$ follow. The fitted per-grid parameters reproduce the statistics of genuine drift, and sampling from the fitted model yields realistic synthetic ENF traces for training (Section IV-B, Fig. 1).

$$dX_t = \theta\,(\mu - X_t)\,dt + \sigma\,dW_t \tag{2}$$

$$X_{t+\Delta} = \mu + e^{-\theta\Delta}(X_t - \mu) + \varepsilon_t, \quad \varepsilon_t \sim \mathcal{N}\left(0, \frac{\sigma^2}{2\theta}\left(1 - e^{-2\theta\Delta}\right)\right) \tag{3}$$

$$Y_t = X_t - \mu, \quad \hat{b} = \frac{\sum_t Y_t\, Y_{t+1}}{\sum_t Y_t^2}, \quad \hat{\theta} = -\frac{\ln \hat{b}}{\Delta}, \quad \sigma_\infty = \frac{\sigma}{\sqrt{2\theta}} \tag{4}$$

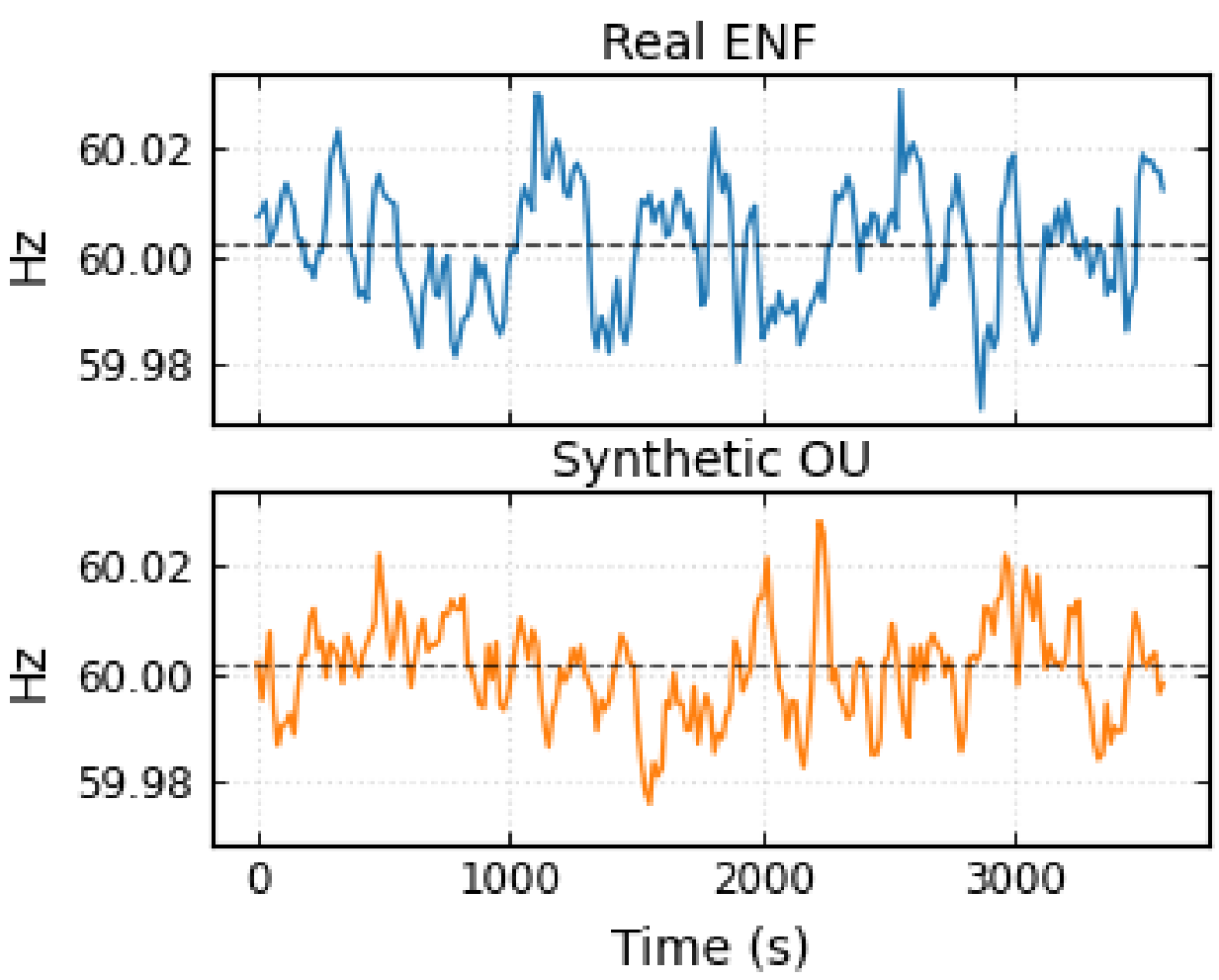

Fig. 1. Measured grid drift (grid A of the MAST dataset [23]) against synthetic drift from the fitted Ornstein-Uhlenbeck model for a single example trace.

### C. Two axes of authenticity

Two distinct questions are not always distinguished in video forensics. The first is whether the footage was captured from a real scene at all, its recording authenticity; the second is whether the depicted content has been manipulated, for example by swapping a face into otherwise genuine footage. ENF speaks to the first: a coherent, reference-consistent ENF is strong evidence that a real scene was recorded under mains lighting. It says little about the second, because a genuine recording can be edited after capture. The framework therefore treats the two as separate axes (Section IV-F), and an ENF-authentic recording is not assumed to be content-authentic.

A further asymmetry governs the ENF axis itself: presence of a coherent ENF is strong positive evidence, but absence is weak, because many genuine recordings carry no usable ENF. A professional audio chain rejects mains hum, and a clip that is dimly lit, shot outdoors, or captured at a degenerate frame rate may yield no reliable flicker. The orchestrator is built around this asymmetry rather than treating absence as incriminating.

### D. Threat model

The adversary may produce fully synthetic video from a text or image prompt, swap or reenact a face within real footage, splice segments from different sources, or misrepresent where and when a genuine clip was captured. The assumption that makes ENF useful is that the adversary cannot reproduce the live, region-specific ENF trace of a claimed time and place: a generator does not measure a live grid (Section VI-B), and an editor of real footage does not control the mains drift already present. Defence against an adversary who splices in a genuinely recorded ENF from the correct grid and time is out of scope, and is noted as such in Section VII.

## IV. The Agentic ENF Framework

The framework couples three electric-network-frequency pathways with four content baselines under a two-axis orchestrator (Fig. 2). The two discrimination channels, audio and optical, share one extractor and one network and judge whether a recording carries a coherent, naturally drifting ENF. A third, authentication channel matches that ENF against an independent grid reference to verify or recover a time of recording. The four content baselines address manipulation of the depicted scene rather than the recording, and the orchestrator fuses all of them along two axes.

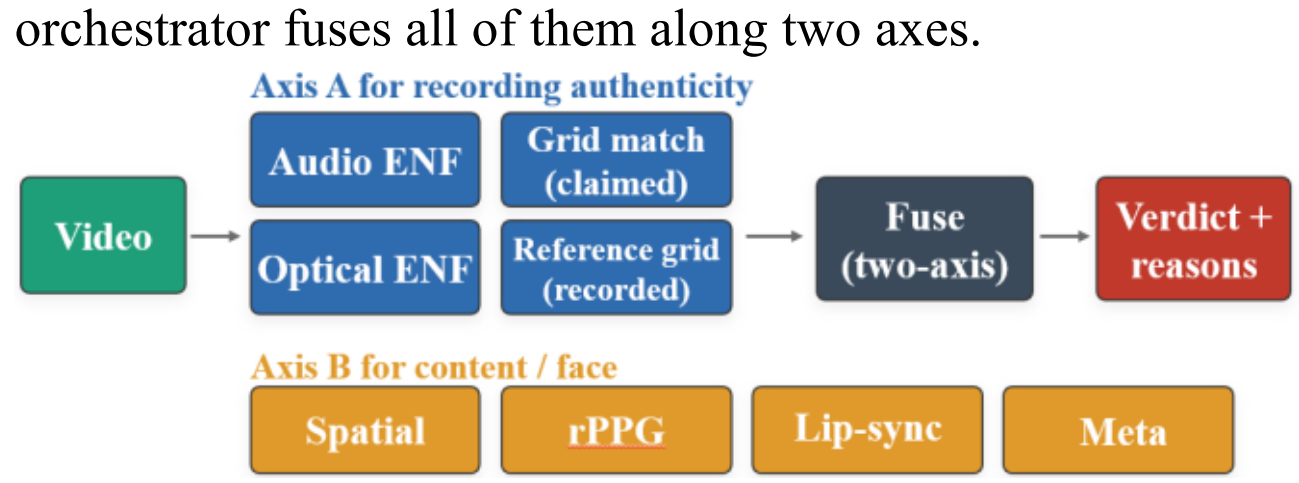

Fig. 2. Overview of the two-axis agentic framework. Axis A judges recording authenticity from audio and optical ENF against a claimed or recovered grid reference, Axis B contributes content and face baselines, and the orchestrator fuses the two and escalates only when the evidence is inconclusive.

### A. Unified two-channel ENF extraction

The estimator band-passes the signal around the second harmonic, so the fundamental ENF is recovered as $\hat{f}_0 = \hat{f}_h / h$ (5), where $\hat{f}_h$ is the tracked frequency of the h-th harmonic and h = 2. It then computes a short-time Fourier transform, locates the dominant peak, and refines it to sub-bin accuracy by parabolic interpolation (6): α, β and γ are the magnitudes of the bin below the peak, the peak bin, and the bin above; δ is the resulting fractional-bin offset; and f_k and Δf are the peak-bin frequency and the bin spacing, so the refined harmonic estimate is $\hat{f}_h$ = f_k + δ Δf. For audio the input is the waveform; for video it is the per-row luminance of a rolling-shutter sensor, which is read out far faster than the frame rate and so resolves the 100 or 120 Hz flicker.

$$\hat{f}_0 = \frac{\hat{f}_h}{h} \tag{5}$$

$$\delta = \frac{1}{2}\frac{\alpha - \gamma}{\alpha - 2\beta + \gamma}, \quad \hat{f}_h = f_k + \delta\Delta f \tag{6}$$

The optical front end requires two corrections, both validated against a simultaneously recorded power reference at a mean correlation of 0.98 across clips within the operating envelope (Fig. 3). First, the temporal mean of each row is removed to suppress the static scene, which would otherwise dominate at the frame rate. Second, the one-dimensional carrier is reconstructed on the true rolling-shutter timeline, placing each row at its real exposure instant t(n, r) = n/f_ps + r·T_ro/H (7), where n is the frame index, r the row index, f_ps the frame rate, T_ro the rolling-shutter readout time, and H the number of rows; the inter-frame idle period, during which no row is exposed, is zero-padded. Without these two corrections the recovered line is an artefact rather than the ENF.

$$t(n,r)=\frac{n}{f_{\mathrm{ps}}}+\frac{r\,T_{ro}}{H} \qquad (7)$$

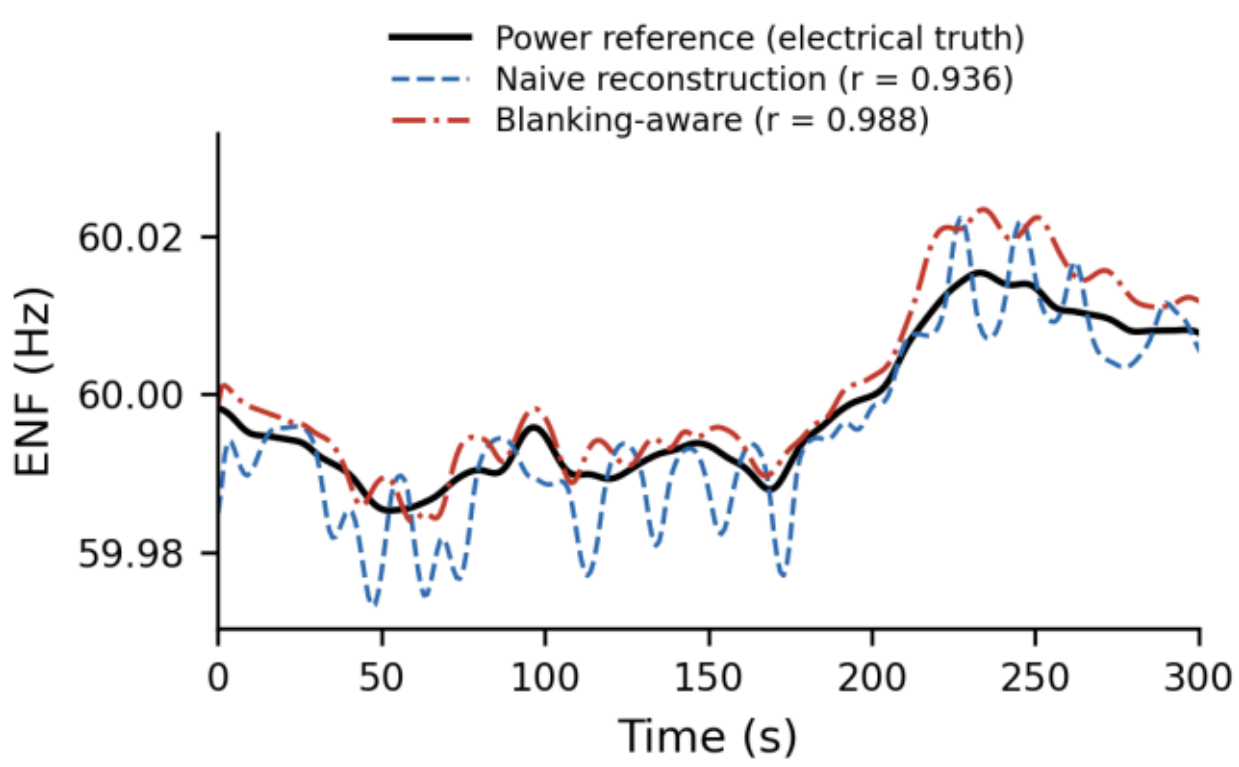

Fig. 3. Optical front end. The second-harmonic flicker recovered from a rolling-shutter video under naive and blanking-aware reconstruction, compared with the simultaneous electrical reference. Both reconstructions have the static scene removed; they differ only in whether the rolling-shutter blanking is modelled.

Recovery has an operating envelope. Below about 20 frames per second the flicker aliases (at 16 fps the 120 Hz line folds to roughly 112 Hz), and at the standard capture rates a frame-rate harmonic can fall in the measurement band and be mistaken for the grid; this degeneracy is characterised separately [8]. A single-clip nominal grid read is therefore treated as unreliable, and the system relies on drift structure and reference matching instead.

## *B. Synthetic-injection training*

Labelled ENF data are scarce, so the detector is trained on synthesised examples anchored to a small quantity of real recordings. A grid's drift is modelled as an Ornstein-Uhlenbeck process fitted to real mains measurements, and a continuous ENF drawn from it is injected into a hum-free carrier: clean speech for the audio channel, and rolling-shutter video whose own flicker has first been suppressed for the optical channel. The optical injection places the modulation at each row's true exposure time, so extraction inverts the same geometry that generated it (Fig. 4, Table I).

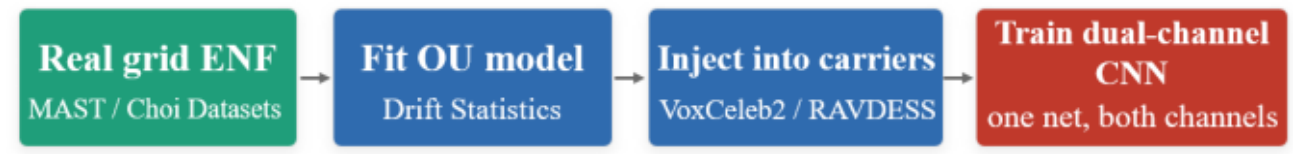

Fig. 4. Synthetic-injection training. Grid ENF is fitted with an Ornstein-Uhlenbeck model and injected into hum-free carriers to train one dual-channel network.

TABLE I
TRAINING-DATA COMPOSITION BY CHANNEL

| Channel | ENF carrier | Real anchor | Classes | Real fraction |
|---|---|---|---|---|
| Audio | VoxCeleb2 [24], RAVDESS [26] | MAST recordings [23] | 6 (genuine + 5 synthetic) | 0.5 |
| Optical | Choi video (flicker suppressed) [25] | Choi power reference [25] | 6 (genuine + 5 synthetic) | 0.5 |

Six classes cover the main ways an ENF can be present, absent or inconsistent: a genuine drifting line; an absent line; a flat, non-drifting line; a phase discontinuity; a cross-regional line at the wrong nominal frequency; and a splice of two segments. Classes are sampled uniformly at training time, so every example is freshly synthesised and the stream is balanced by construction. The genuine class is split evenly between real anchor windows from held-in recordings and synthetic carriers with an injected genuine drift; the governing rule is that real recordings appear only as the genuine class, so every fake is synthetic and separation is never learned from real fakes (Fig. 5).

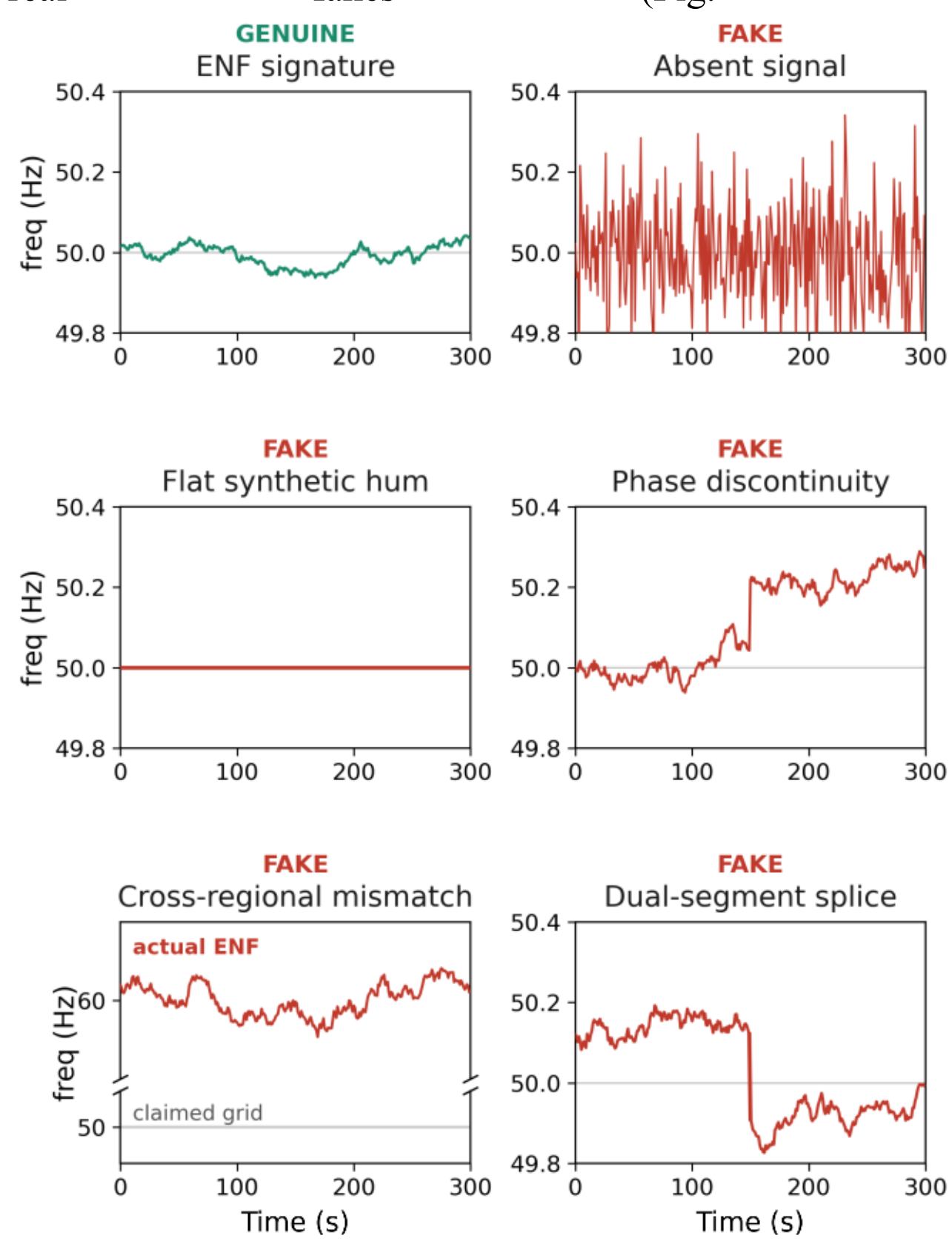

Fig. 5. The genuine ENF signature and the five synthetic manipulation classes used for injection training.

## C. The ENF detector

Both channels use the same network. Two convolutional branches read the ENF spectrogram, a fully connected branch reads summary statistics of the instantaneous-frequency trace (peak prominence, drift magnitude and variability), and the branches are fused for a per-class decision. Adaptive pooling makes the network invariant to input size, so the audio and optical features, which differ in length, feed the same architecture unchanged. The model is deliberately small, ≈ 157,600 parameters, which suits the limited training signal and keeps inference cheap enough to run inside the orchestrator (Fig. 6).

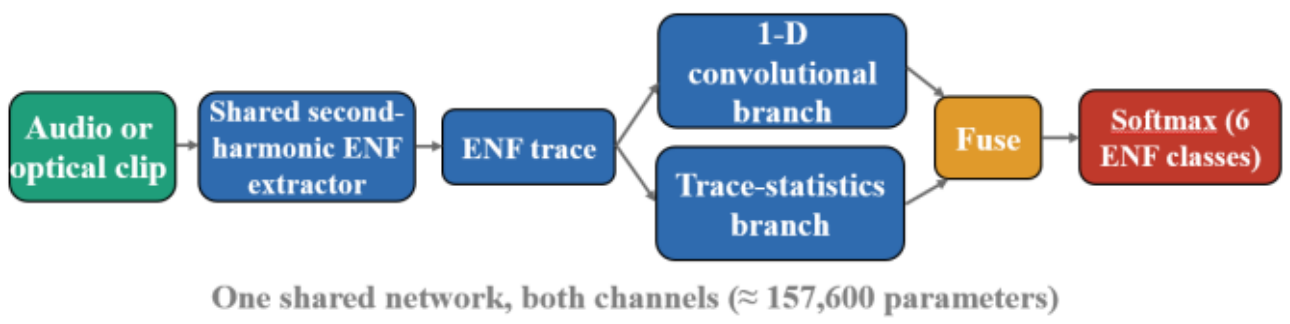

Fig. 6. The unified ENFDetector. One shared second-harmonic extractor feeds a 1-D convolutional branch and a trace-statistics branch, fused into a six-class head, and serves both the audio and optical channels.

## D. Reference-matching authentication

Presence of an ENF is weak evidence, because generated video carries an ENF-band line too; consistency with an independent grid reference is strong evidence, because a generator does not measure a live grid. The authentication channel therefore compares a recording's ENF against a reference trace from a grid database. Instantaneous frequency is extracted by the analytic (Hilbert) method rather than the short-time transform, which raises the fraction of recordings that yield a usable trace from 10% to 38%. Two operations are supported. Verification scores the match at a claimed time as the normalised cross-correlation R_m(τ) between the recording's ENF x(t) and the reference r(t) at the lag τ implied by that time (8), where x̄ and r̄τ are the means of the recording trace and of the lagged reference window. Blind search instead slides the reference over all lags and takes the peak, S_ref = maxτ R_m(τ) (9), recovering an unknown time of recording. The channel emits the same result type as the detectors and enters the orchestrator unchanged.

$$R_m(\tau) = \frac{\sum_t (x(t) - \bar{x})(r(t+\tau) - \bar{r}_\tau)}{\sqrt{\sum_t (x(t) - \bar{x})^2 \sum_t (r(t+\tau) - \bar{r}_\tau)^2}} \quad (8)$$

$$S_r ef = \max_\tau R_m(\tau) \quad (9)$$

## E. Content baselines

The content axis is covered by four decoupled detectors behind a common interface. A spatial detector applies a pretrained deepfake classifier to sampled frames; an rPPG detector tests for a physiologically plausible pulse; a lip-sync detector measures the correspondence between mouth motion and the audio envelope; and a metadata detector assesses provenance and the credibility of any claimed grid region. The rPPG and lip-sync detectors are self-referential: each scores its cue against the clip's own null, a frame-shuffle or an audio roll, so a present cue is evidence while an absent one is uninformative. Each baseline returns the common result type and is consumed by the orchestrator with no change to its logic.

## F. Two-axis agentic orchestrator

The detectors are fused along two axes. Axis A, the ENF pathways, judges recording authenticity, whether the footage was captured from a real scene under mains lighting. Axis B, the content baselines, judges manipulation of the depicted content. The axes are asymmetric, and the fusion rules reflect this: an ENF present in either modality is strong evidence of a genuine recording, so the orchestrator never requires both channels to agree, whereas a confident content-level fake overrides an ENF-genuine verdict, because a real scene can be filmed and then have its content swapped. This override resolves the face-swap-on-real-footage case that a naive conjunction would miss (Fig. 2).

Fusion is expressed as a deterministic decision graph (Fig. 7) and implemented both directly and as an equivalent state machine, the two verified to agree. To bound cost, the graph runs as an escalating cascade: the cheap, high-weight detectors run first, and the expensive, face-based rPPG and lip-sync detectors are invoked only when that tier is inconclusive and a face is present. Section VI shows that this cascade matches the accuracy of evaluating every detector while cutting wall-clock time by 54%.

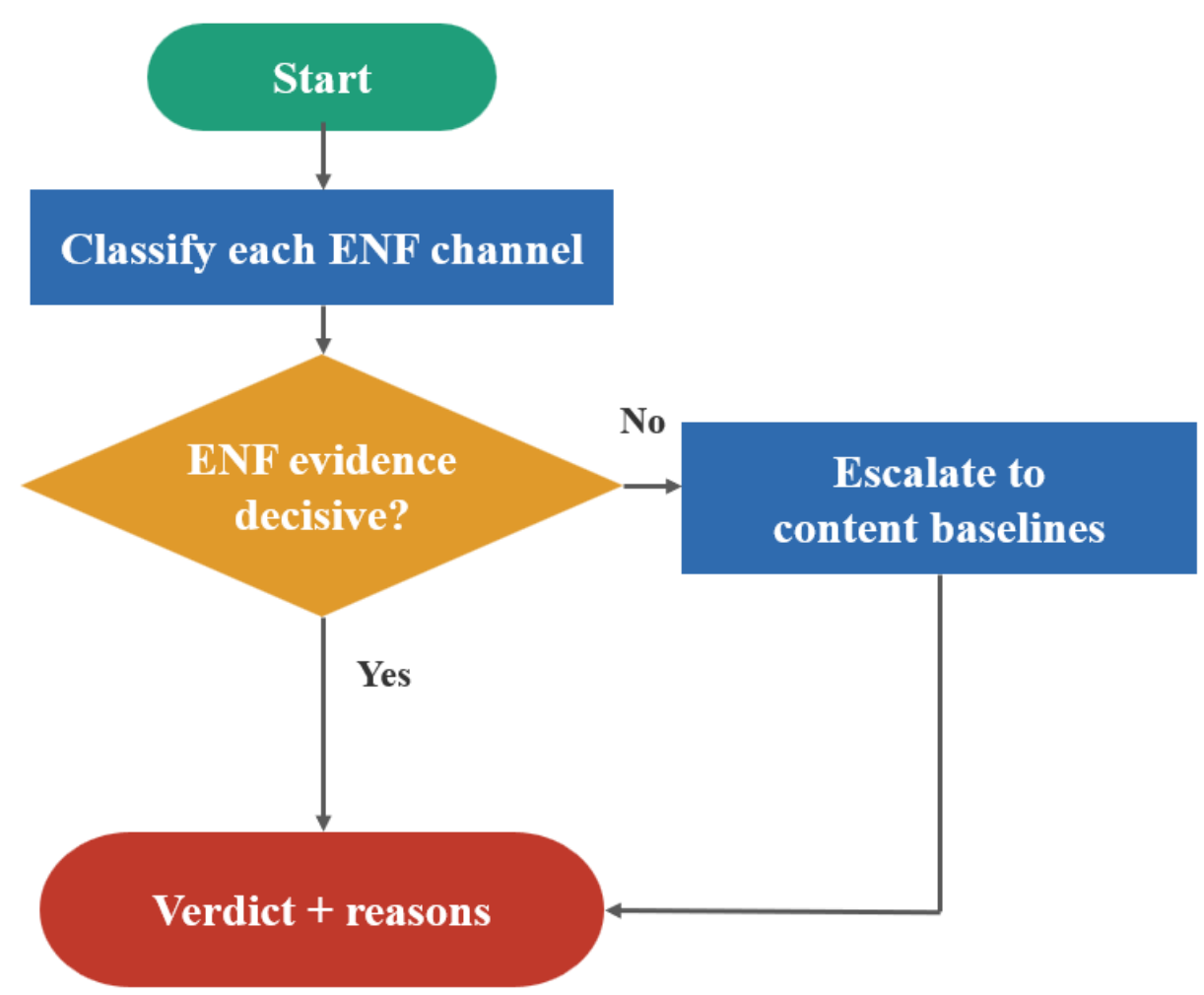

Fig. 7. The agentic orchestrator as a deterministic state graph: assess the ENF evidence, apply the fusion rules, and escalate to the content baselines only when the ENF verdict is inconclusive.

# V. Experimental Setup

## A. Datasets

Nine sources are used, each in a defined role (Table II). The MAST corpus [23], which pairs audio with a simultaneous power reference at 1 kHz, calibrates the drift model and provides the real audio-ENF anchor. VoxCeleb2 speech [24], verified to carry no coherent ENF, is the hum-free audio carrier. The Choi rolling-shutter videos [25], each with a simultaneous power recording, provide optical ground truth and the real optical-ENF anchor; the two clips shot at 16 fps are excluded because the flicker aliases below the operating envelope. RAVDESS supplies the optical injection carrier, its own strong flicker suppressed before a controlled ENF is injected [26]. The ENF-WHU corpus [27] supports claimed-time audio authentication, and a real scene with its grid reference [28], together with two AI recreations of the same scene, supports the optical authentication study. Output from six text-to-video generators, sampled from VBench-2.0 [1], forms the generated set, and FakeAVCeleb [29], which contains real face-swaps, is held back as the out-of-distribution test. The MAVOS-DD corpus [30], [31] calibrates the detection ensemble and serves as a frozen, by-language test set.

TABLE II
DATASETS AND THEIR ROLES

| Dataset | Role | Access | Scale used |
|---|---|---|---|
| MAST [23] | Drift-model fit; real audio-ENF anchor and in-distribution audio test | IEEE DataPort | 9 grids (A–I), 1 kHz, 30–60 min each |
| VoxCeleb2 [24] | Hum-free audio carrier for injection | Public (Kaggle) | 5,000 carriers from a 151k-clip pool |
| RAVDESS [26] | Optical carrier (flicker suppressed, then injected) | Public (Zenodo) | 1,533 of 4,902 clips |
| Choi [25] | Optical round-trip ground truth; real optical-ENF anchor | IEEE DataPort | 11 videos, each with a synchronous power recording |
| ENF-WHU [27] | Claimed-time audio authentication reference | Public (GitHub) | 130 recordings + 24 h reference |
| EV-ENFD [28] | Optical authentication: rolling-shutter video + grid reference | Public | 91 recordings + grid reference |
| MAVOS-DD [30], [31] | Ensemble calibration and frozen test (by language) | Public (Kaggle) | 9,600 clips (4 languages, 4,800 real / 4,800 fake) |
| FakeAVCeleb [29] | Out-of-distribution real face-swaps for the ensemble | On request | 40 within a 90-clip out-of-distribution set |
| VBench-2.0 [1] | Cross-generator ENF survey; fully synthetic class | Public (Hugging Face) | 6 generators, 100 clips each (600) |

### B. Evaluation protocol

One rule governs the data: real recordings appear only as the genuine class, and all manipulated classes are synthetic. Separation of genuine from manipulated is therefore measured on synthetic data, where the labels are exact, while acceptance of genuine material is measured on held-out real recordings from sources unseen in training. Splits are made by source to prevent leakage: the MAST recordings and the Choi videos are each partitioned into a held-in anchor set and a disjoint held-out test set, by recording and by video respectively. Two test sets are used per channel: a frozen synthetic evaluation over all six classes for separation, and a genuine-only held-out real set for acceptance. FakeAVCeleb, evaluated on the full ensemble, is the first test that contains real manipulations and unseen capture conditions. The ensemble threshold is calibrated and frozen on a multilingual deepfake set (MAVOS-DD [30], accessed through a restructured Kaggle release [31]), split by language so that calibration and test never share a language.

### C. Metrics and implementation

Discrimination is reported as accuracy, macro-averaged F1 and the area under the receiver operating characteristic (ROC) curve (AUC); genuine-acceptance as the proportion of held-out real recordings accepted; and the ensemble study additionally as the false-accusation rate and, for the agentic comparison, wall-clock time and the number of frames decoded. Paired comparisons against the naive baselines use McNemar's exact test. The system runs in Python with PyTorch on a single consumer graphics processing unit, and the orchestrator is realised both as direct logic and as an equivalent state machine.

## VI. RESULTS

### A. ENF discrimination and real-data transfer

On held-out real recordings from unseen sources, the detector genuine-acceptance rate is 93% for audio and 100% for optical, against naive presence baselines of 53% and 15%; the synthetic-injection training with a real anchor is what closes this transfer gap. On the binary genuine-versus-manipulated task the detector reaches 99% accuracy for audio and 89.4% for optical, against naive baselines of 63% and 52.5%, both improvements significant under McNemar's exact test ($p < 10^{-27}$). On the frozen synthetic evaluation over all six classes the audio channel attains 99.0% accuracy (macro-F1 0.990, AUC 0.9995) and the optical channel 88.3% (macro-F1 0.871), the structural fakes recognised at close to 100% and the error concentrated in the genuine-versus-flat pair (Table III, Figs. 8 and 9).

TABLE III
HEADLINE RESULTS BY COMPONENT

| Component | Measure | Result |
|---|---|---|
| Audio ENF detector | Genuine-acceptance / six-class accuracy | 92.9% / 99% |
| Optical ENF detector | Genuine-acceptance / six-class accuracy | 100% / 88.3% |
| Audio detector (synthetic) | Macro-F1 / AUC | 0.99 / ≈1.0 |
| Audio authentication | Claimed-time AUC | 1.0 (0.774 over all 40) |
| Audio authentication | Blind timestamp recovery | ≈1 s in 24-h reference |
| Optical authentication | Cross-modal agreement | ≈0.80 |
| Cross-generator survey | Flicker separates real vs synthetic AUC | 0.51 (no separation) |
| Consistency procedure | False accusations, 91 genuine recordings | 0 of 91 |
| Agentic cascade | Wall-clock / frames saved vs full ensemble | 54% / 64% |

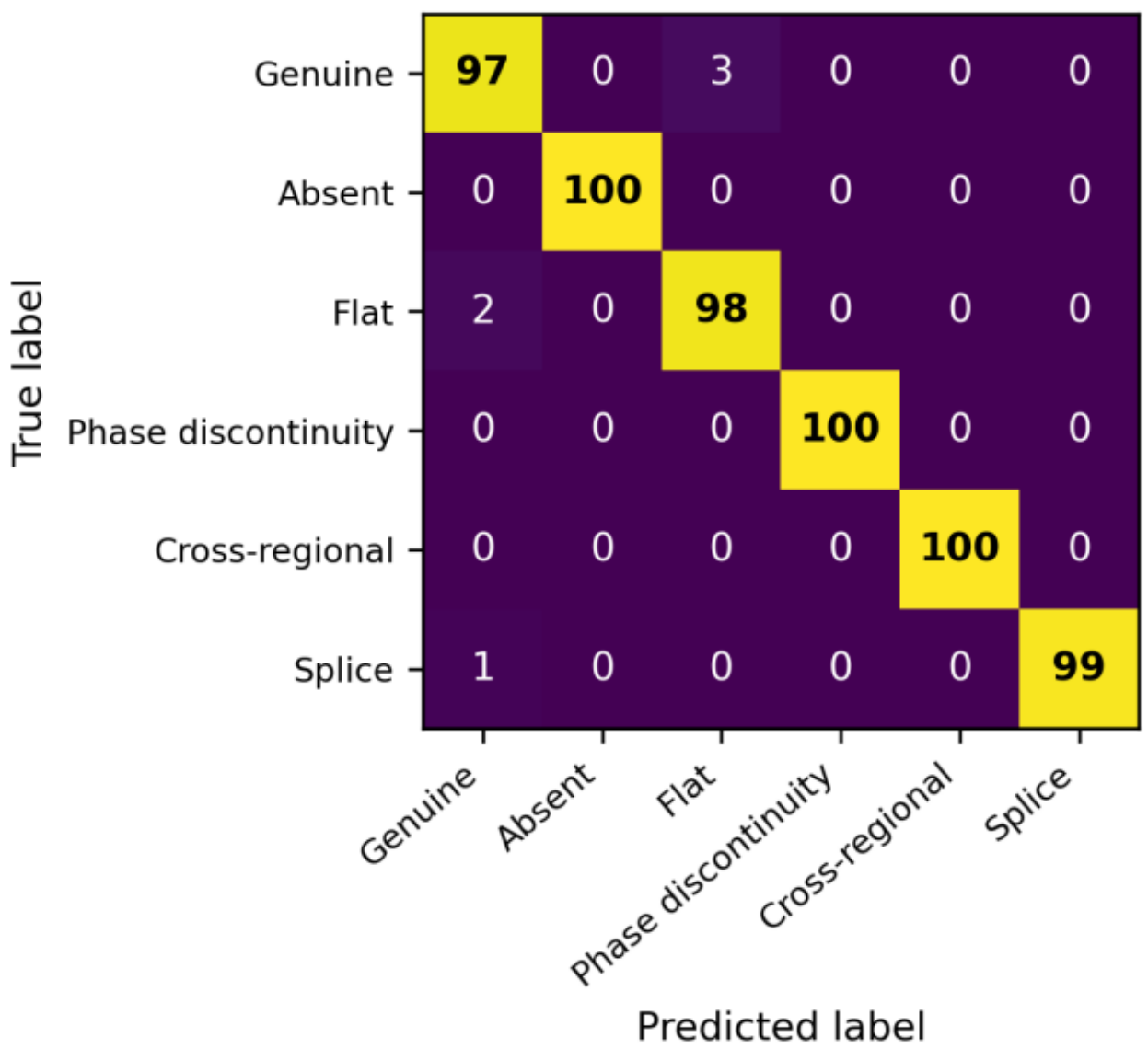

Fig. 8. Confusion matrix for the audio ENF detector on the frozen synthetic six-class evaluation.

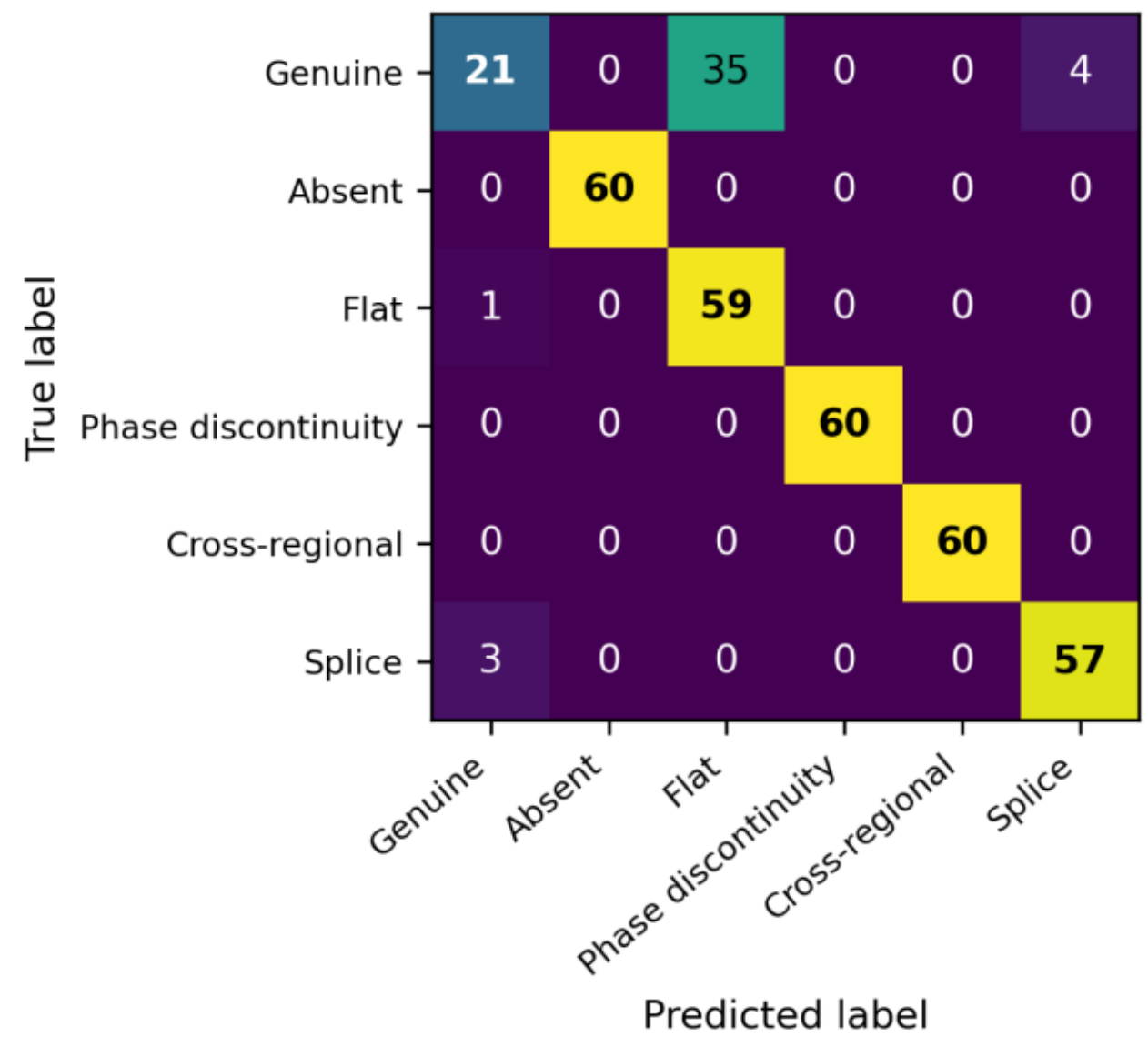

Fig. 9. Confusion matrix for the optical ENF detector on the frozen synthetic six-class evaluation.

### *B. Generated video carries ENF flicker*

The premise that synthesis could be flagged by a missing ENF does not hold. Across the generated set the optical flicker is present at a prominence comparable to real footage, so the strength of an in-band line does not separate genuine from generated video, and each generator's apparent grid is set by its characteristic frame rate rather than measured from a live supply. Even the nominal grid that a single short clip appears to report is confounded by the camera frame rate; this confound is characterised separately [8]. This finding motivates the shift from detection by absence to reference-matched authentication (Fig. 10).

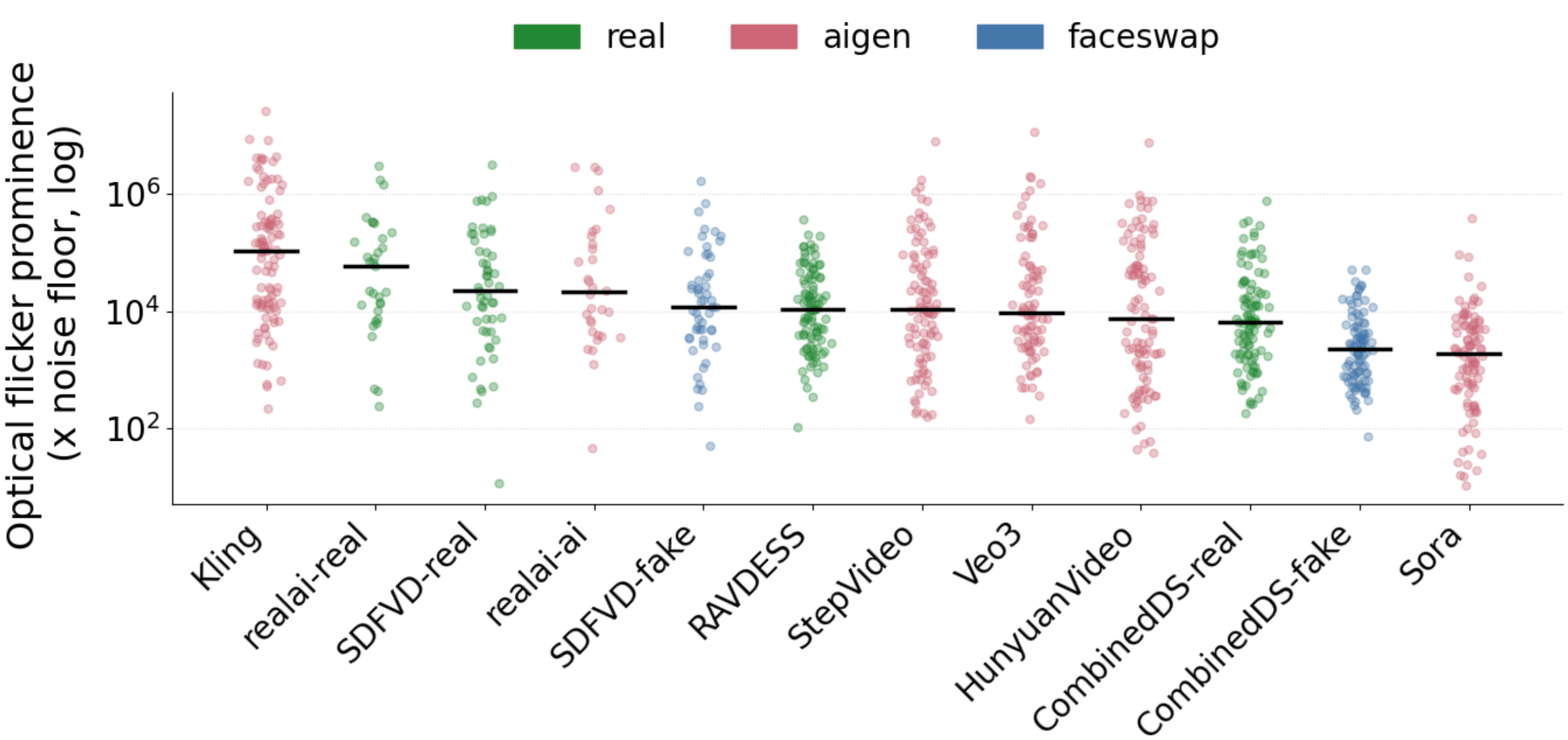

Fig. 10. Optical flicker prominence across real, generated and face-swap sources, as a multiple of the noise floor (log scale). The generated clips are not separable from the real ones (AUC 0.51).

*C. Reference-matched authentication*

Matching a recording's ENF against an independent grid reference authenticates it where presence alone cannot. On the audio corpus, the analytic-signal extractor raises usable coverage from 10% to 38% over the short-time transform (Fig. 11); on the recordings it recovers, verification of a claimed time separates genuine recordings from false-timestamp and no-ENF cases at an AUC of 1.0, and blind search recovers an unknown recording time to within one second in a twenty-four-hour reference (Fig. 12). On the optical side, a real scene's flicker matches its grid reference and agrees with the same recording's audio ENF (cross-modal correlation 0.80, a reference-free check; Fig. 13), whereas two AI recreations of that scene show a flicker at 60 Hz, the wrong grid for the claimed 50 Hz location, so the recreation carries no live grid from the scene it depicts. A consistency check across the ninety-one genuine EV-ENFD recordings raises no false accusation under the true claimed grid (Fig. 14).

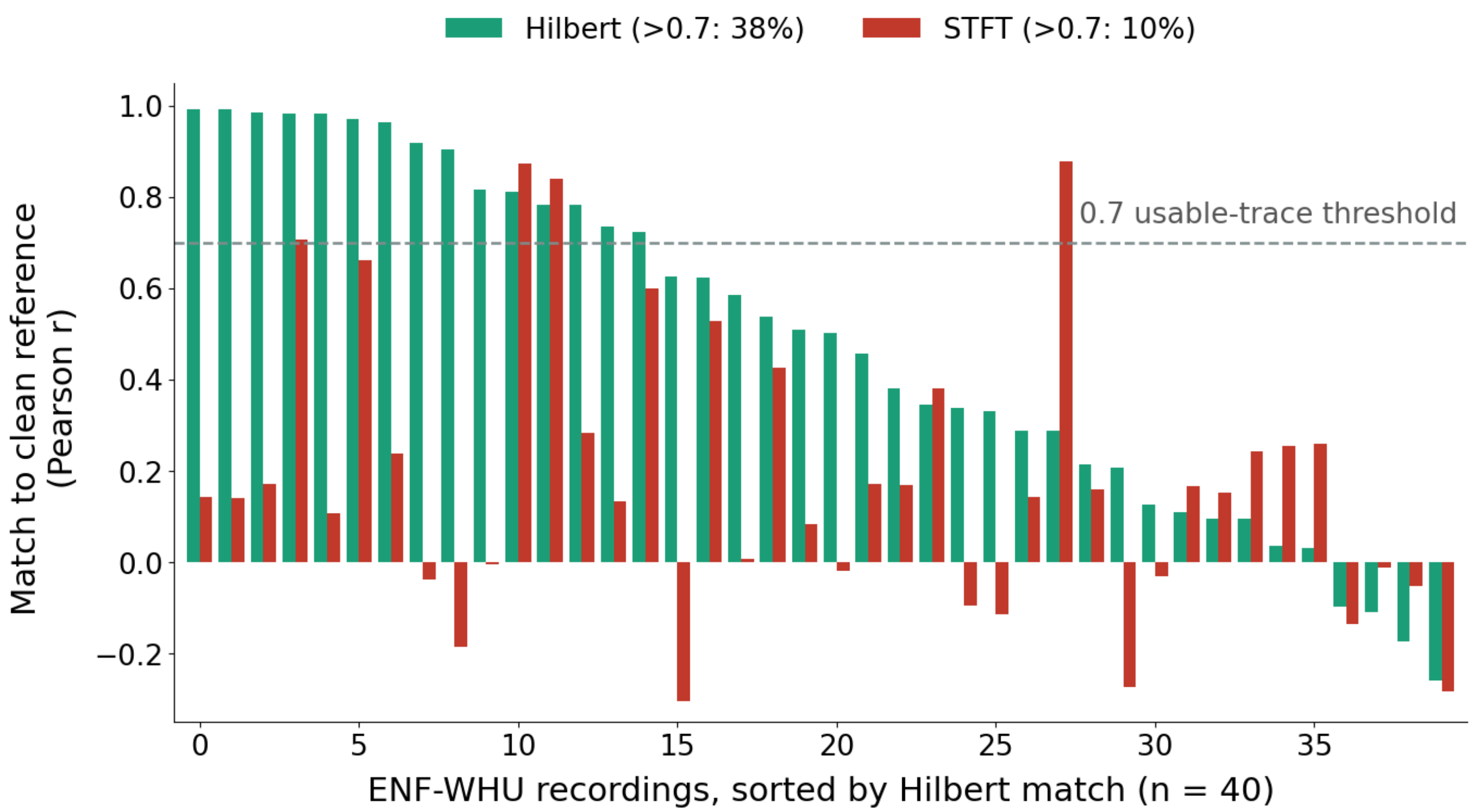

Fig. 11. Authentication coverage. Match to the clean grid reference for each recording, the analytic-signal (Hilbert) extractor recovering a usable trace on far more clips than the short-time transform.

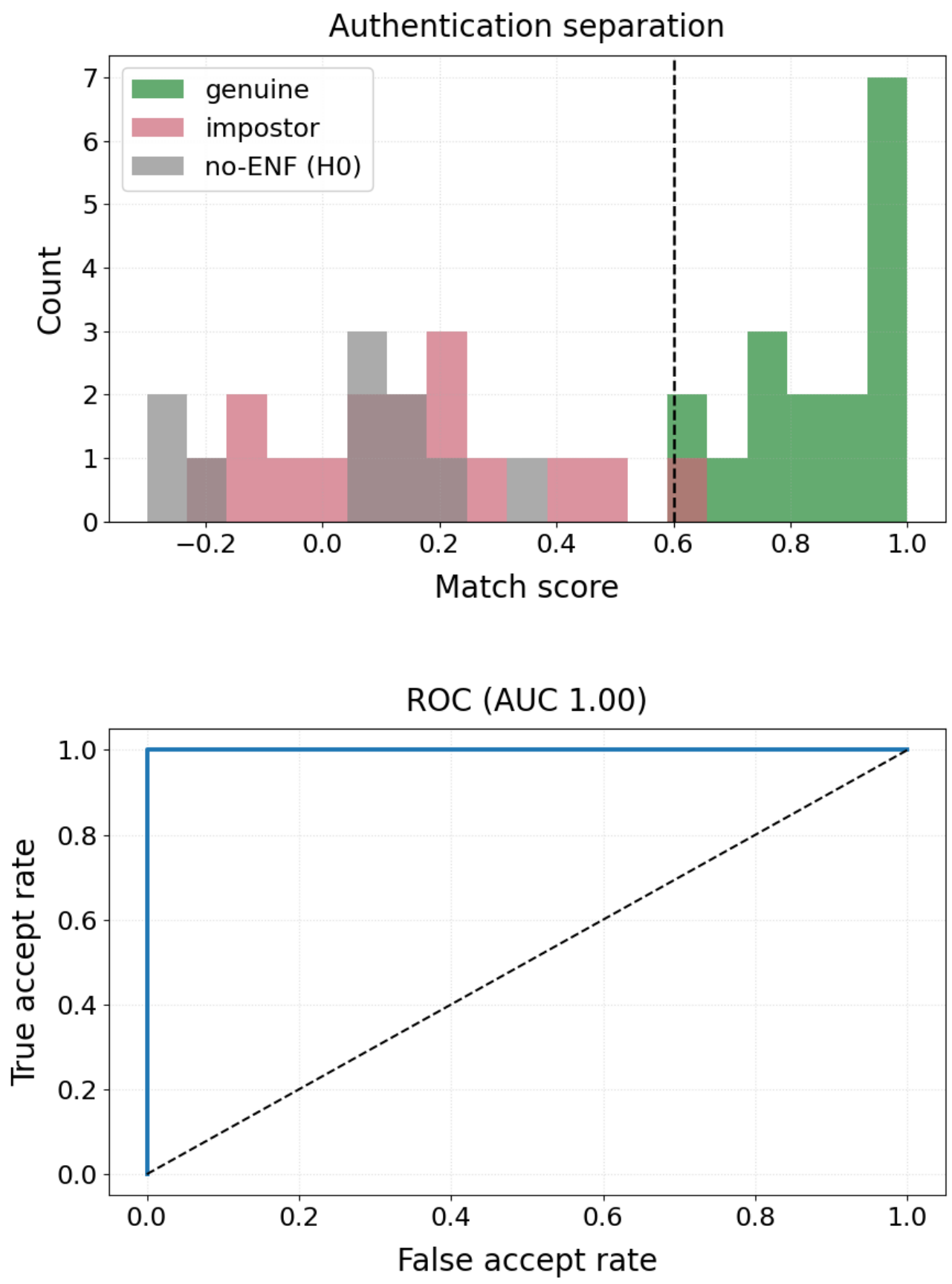

Fig. 12. Claimed-time authentication. Match-score separation between genuine, impostor and no-ENF cases, and the corresponding ROC curve (AUC 1.00).

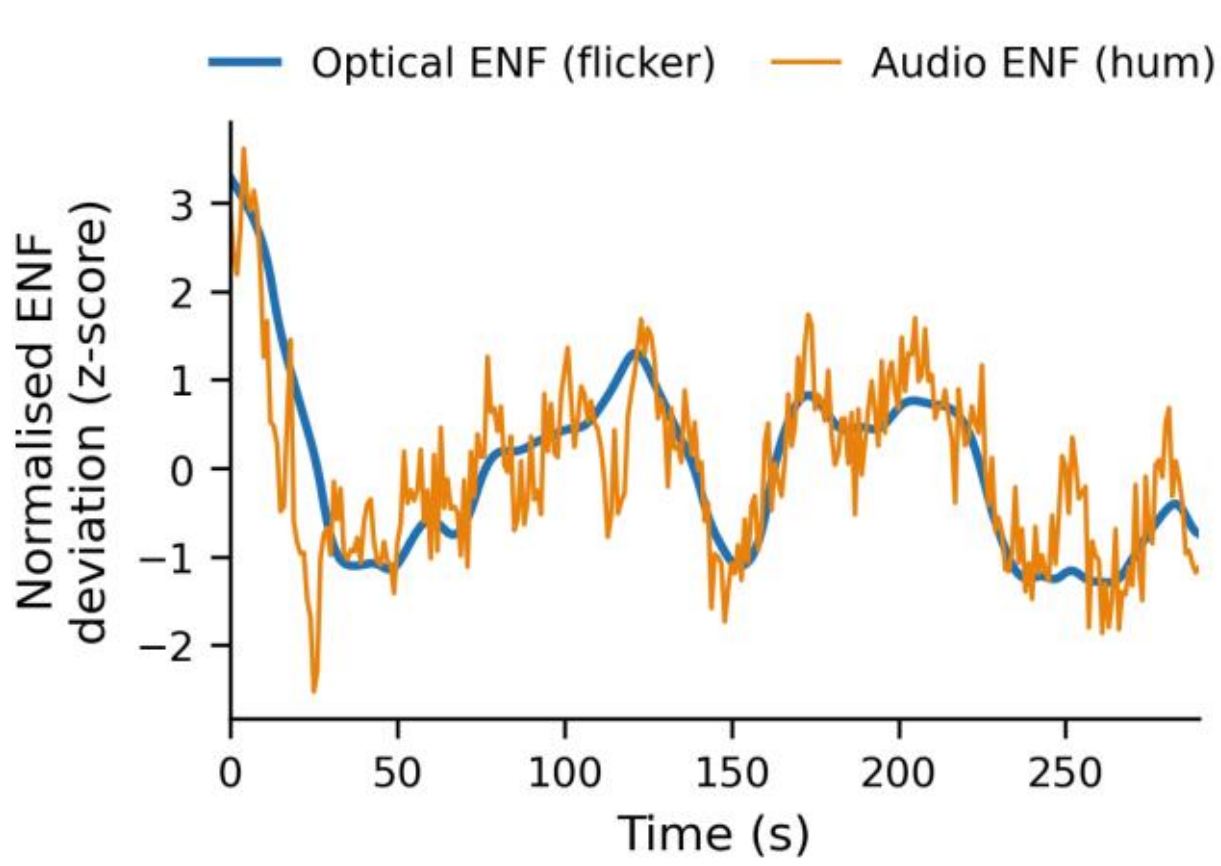

Fig. 13. Reference-free cross-modal consistency: the optical flicker and the audio hum of the same recording follow the same trajectory (correlation 0.80).

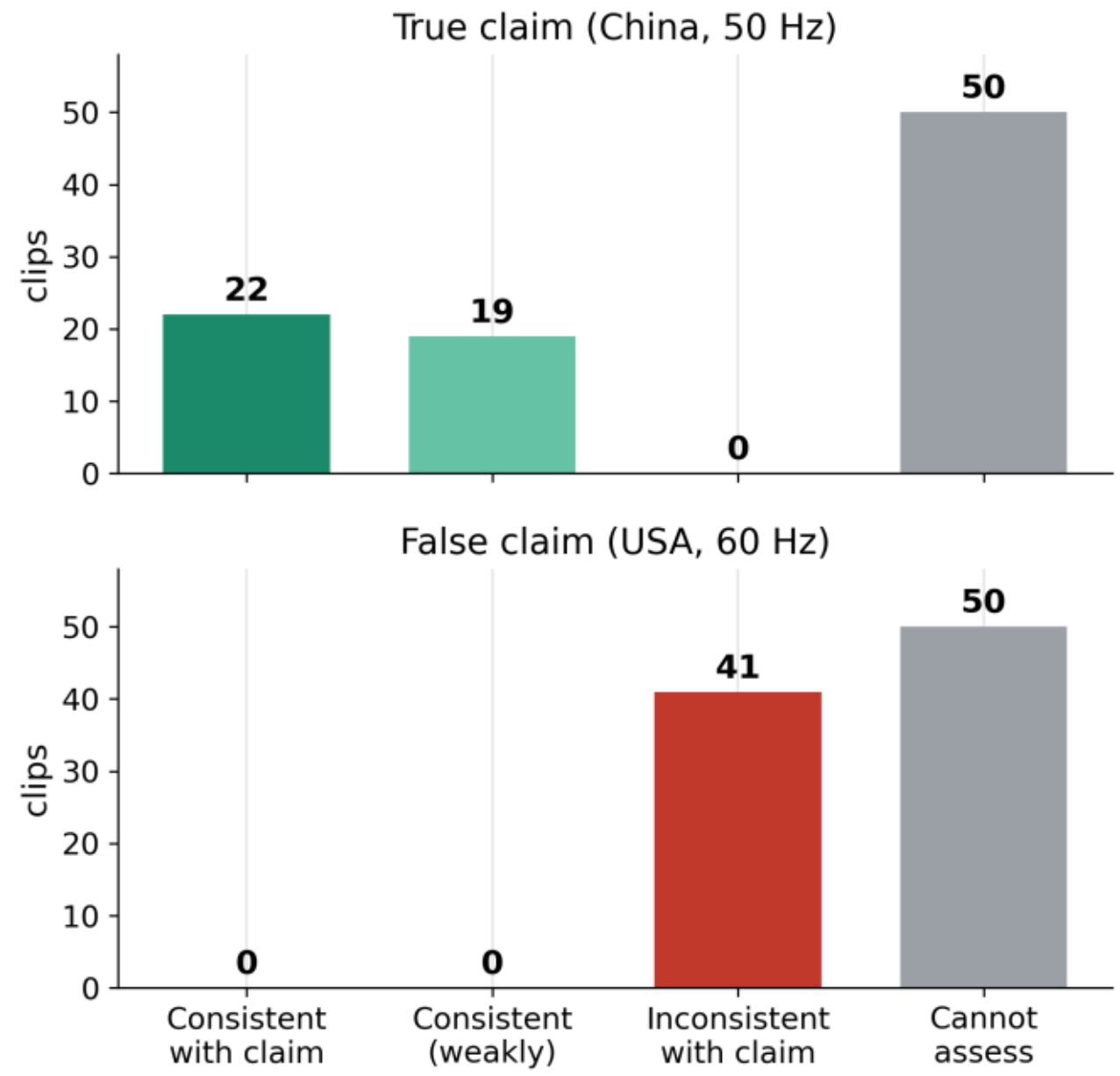

Fig. 14. Consistency verdicts on ninety-one genuine EV-ENFD recordings under a true and a false claimed grid; the false-accusation rate under the true claim is zero.

### *D. The agentic ensemble*

On the out-of-distribution FakeAVCeleb test the two-axis routing and a flat averaging ensemble reach the same accuracy, and routing reduces the false-accusation rate from 0.47 to 0.30 by trusting an ENF-genuine verdict unless a content baseline is confidently fake, although this difference is not significant at this sample size (McNemar, p = 0.27; Fig. 15, Table IV). Run as an escalating cascade, the framework matches the accuracy of evaluating every detector while cutting wall-clock time by 54% and frames decoded by 64%, because the expensive detectors are invoked only when the cheap tier is inconclusive (Fig. 16). The agentic policy therefore matches the full ensemble's accuracy at substantially reduced compute.

TABLE IV
THE AGENTIC ENSEMBLE: ROUTING AND CASCADE

| Metric | Flat averaging | Two-axis routing | Agentic cascade |
|---|---|---|---|
| Accuracy | 0.48 | 0.48 | 0.52 |
| False-accusation rate | 0.47 | 0.30 | 0.37 |
| Fake-recall | 0.45 | 0.37 | 0.45 |
| Wall-clock time per clip | 11.3 s | 11.3 s | 5.1 s |
| Frames decoded | 100% | 100% | 36% |

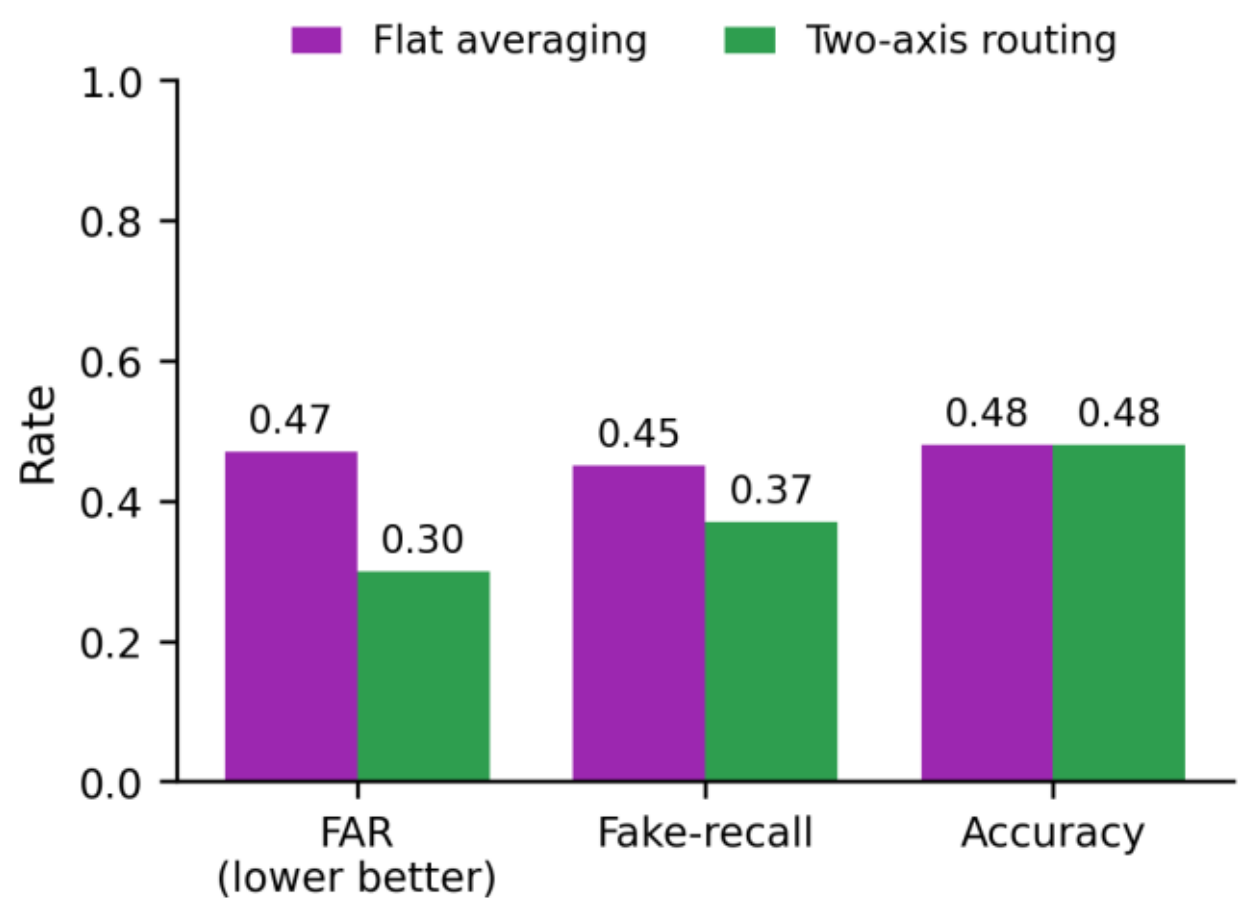

Fig. 15. Accuracy, false-accusation rate and fake-recall for a flat averaging ensemble against the two-axis agentic routing.

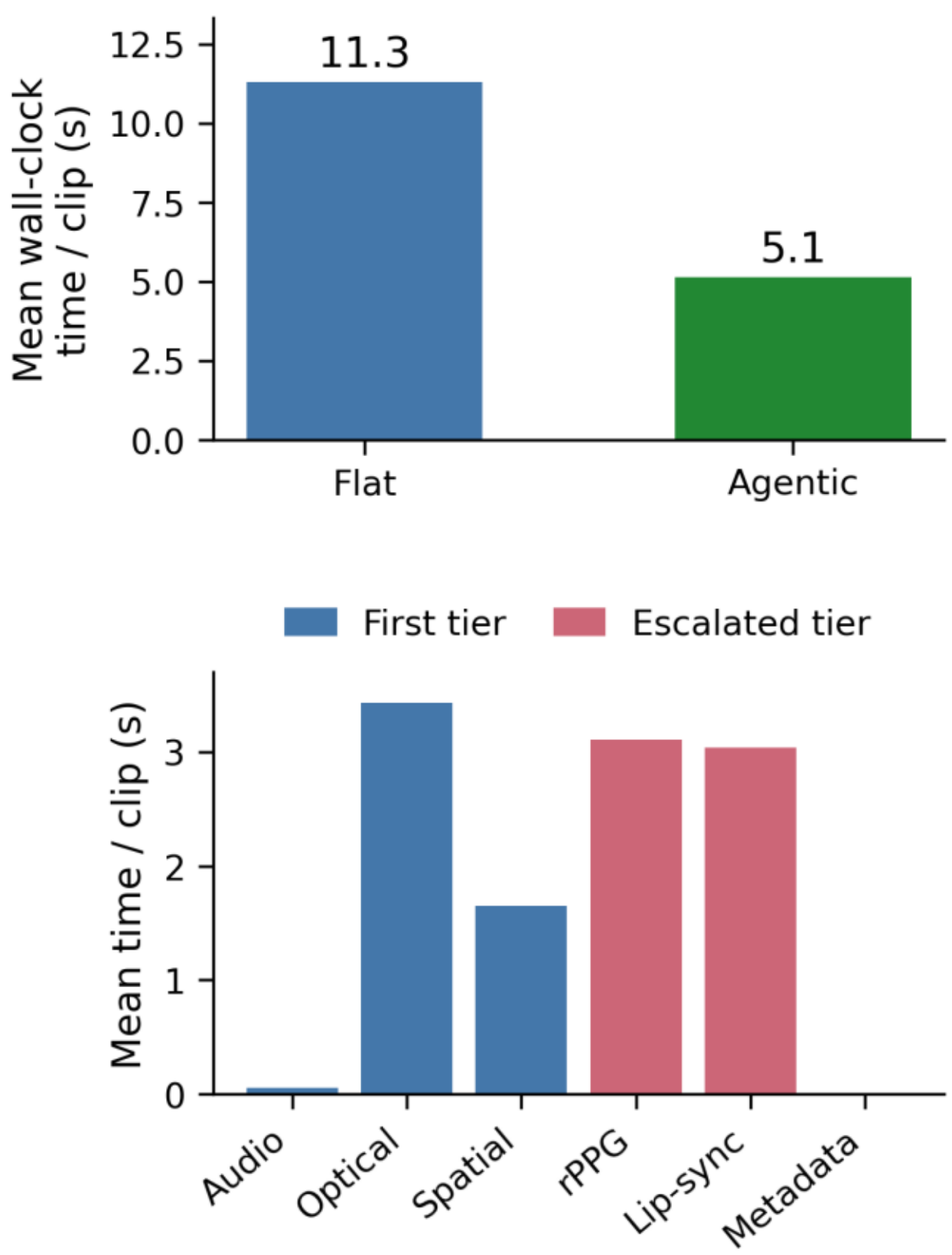

Fig. 16. Computational cost per clip for the full ensemble against the agentic cascade.

*E. Ablations*

Three ablations isolate the decisive components. Removing either optical front-end correction collapses the recovered line into an artefact, confirmed against the simultaneous power reference at a mean correlation of 0.98. Widening the drift jitter of the synthetic genuine class to reproduce the heavy-drift tail of real grids is what lifts held-out optical acceptance to 100%. Replacing the short-time transform with the analytic-signal extractor yields the coverage gain in authentication, and the residual genuine-versus-flat ambiguity is resolved by temporal autocorrelation, which is high for a real ENF and near zero for a flat line (Fig. 17).

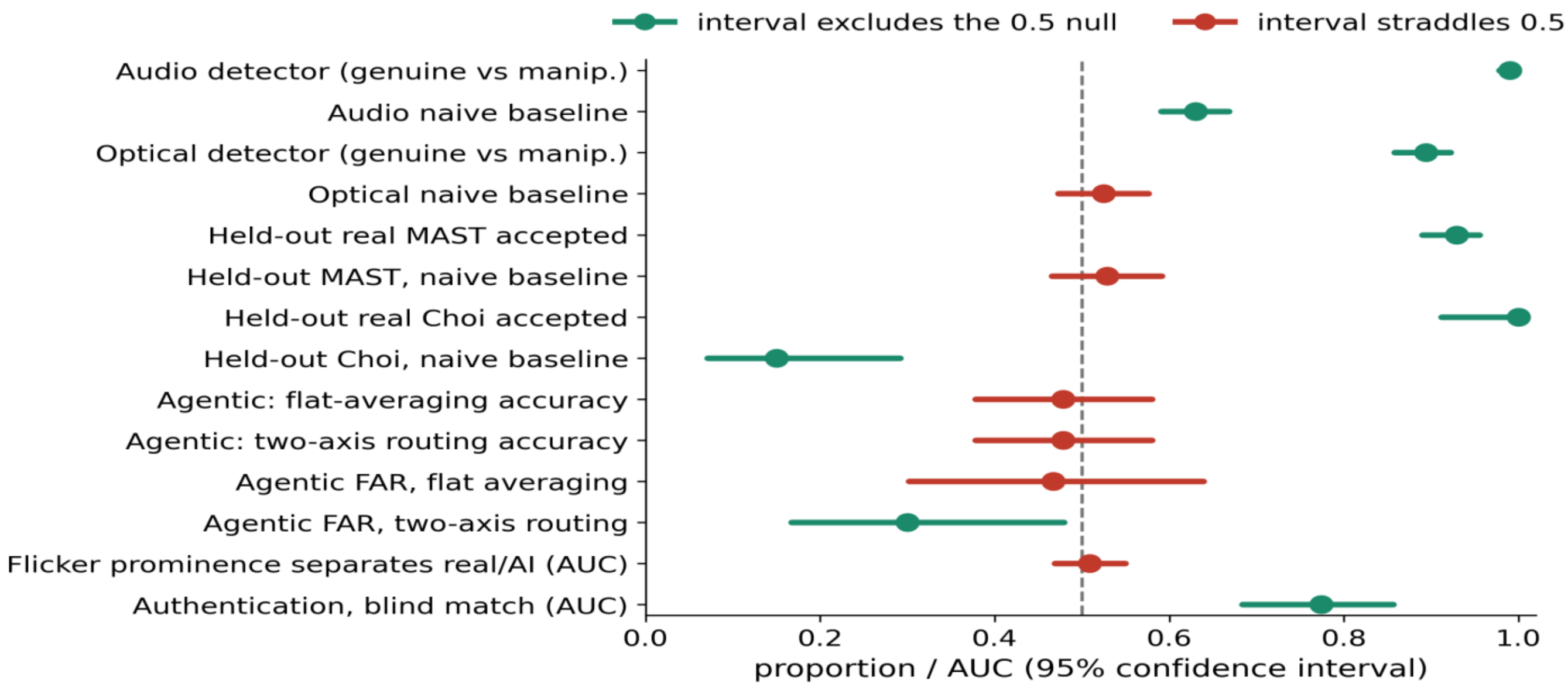

Fig. 17. Every headline claim with its 95% confidence interval; intervals that exclude the 0.5 null are shown apart from those that straddle it.

## VII. Discussion and Limitations

The framework's strengths and its limits both follow from the nature of the signal it relies on. The ENF is a property of the recording rather than of its depicted content: it authenticates the capture but not what the frame shows. The clearest limitation is therefore the face-swap on otherwise genuine footage. Such a clip is itself a genuine recording, so its ENF reads as authentic, and the swapped face must be caught by the content baselines alone. On compressed, out-of-distribution video these perform close to chance, and the orchestrator's override mitigates but does not solve it. This is the acknowledged hard case for an ENF-centred system.

A second limitation concerns ENF presence. A coherent line can be adjudicated from a clip alone only where it is strong; at low signal-to-noise ratio a matched-null test shows that both the mean frequency and the drift shape are confounded, so clip-only presence is not reliable evidence. In-the-wild distribution compounds this: re-encoding and compression by sharing platforms attenuate the ENF and push the signal-to-noise ratio lower still. This is why authentication rests on reference matching rather than presence, and why absence is treated as uninformative rather than incriminating. For the same reason a single short clip cannot report even the nominal grid, which is set by the camera frame rate rather than the supply [8]; the system therefore avoids single-clip region reads and requires a non-degenerate frame rate for optical recovery.

The authentication channel depends on the availability of an independent grid reference. Public records exist for some grids, for example the United Kingdom and continental Europe, while others are access-gated, such as the United States, or unpublished, such as China, which bounds where claimed-time verification can currently be applied. A reference need not come from a public archive, however: the grid signal is recoverable from a simple mains recording, so coverage for any location can be built by capturing it directly, though verifying a past timestamp still needs a reference that was recording at that time. Extending coverage is therefore a matter of reference availability rather than of method.

Finally, the intended use is forensic triage and corroboration, not autonomous judgement. A verdict should be read as one line of evidence, with its per-pathway confidences exposed, and an adversary who splices a genuinely recorded ENF of the correct grid and time into manipulated footage remains out of scope.

## VIII. Conclusion

Once generators reproduce the very signal a detector looks for, authentication is better framed as confirming a genuine recording than as spotting a fake. Built on that reframing, the framework extracts the electric network frequency from both audio and rolling-shutter video with one estimator and one network, trains it from synthetic injections anchored to real recordings, authenticates a claimed or unknown time of recording by matching against an independent grid reference (or, where none is available, by the internal consistency of a clip's own optical flicker and audio hum), and fuses the ENF evidence with content baselines through a two-axis agentic orchestrator. It accepts genuine audio and video at 93% and 100% on held-out real recordings, verifies a claimed time at an AUC of 1.0, and, run as a cascade, matches a full ensemble's accuracy at 54% less wall-clock time and with fewer false accusations, fast enough for forensic triage at scale. Future work includes generalising to real, in-the-wild manipulations, broadening the grid-reference coverage that authentication depends on, and strengthening the content axis that remains the weakest link. The aim is not to declare a clip fake, but to say, with calibrated confidence, when a recording is what it claims to be.

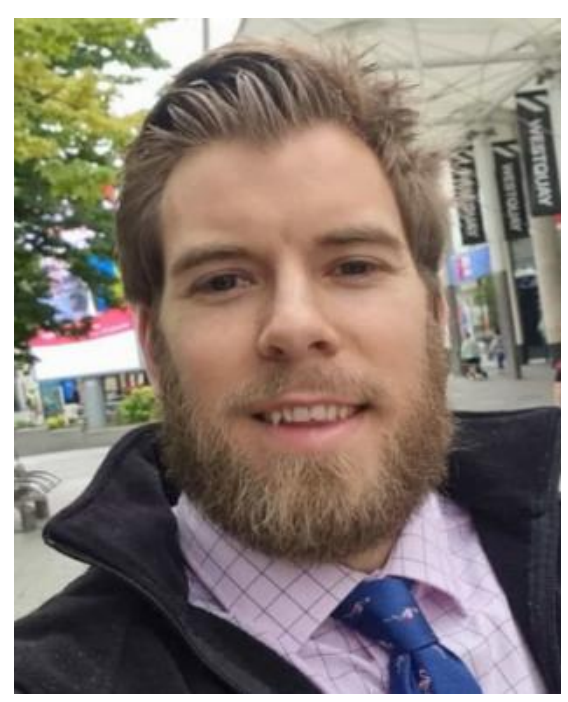

**Ryan Wyton** is an MSc student in Applied AI and Data Science at Southampton Solent University, Southampton, U.K. He received the bachelor's degree in music from the University of Chichester, Chichester, U.K., where he first developed an interest in the analytical properties of acoustic signals, before moving into data science. His research interests include multimedia forensics, the electric network frequency, signal processing, and the detection and authentication of AI-generated media.

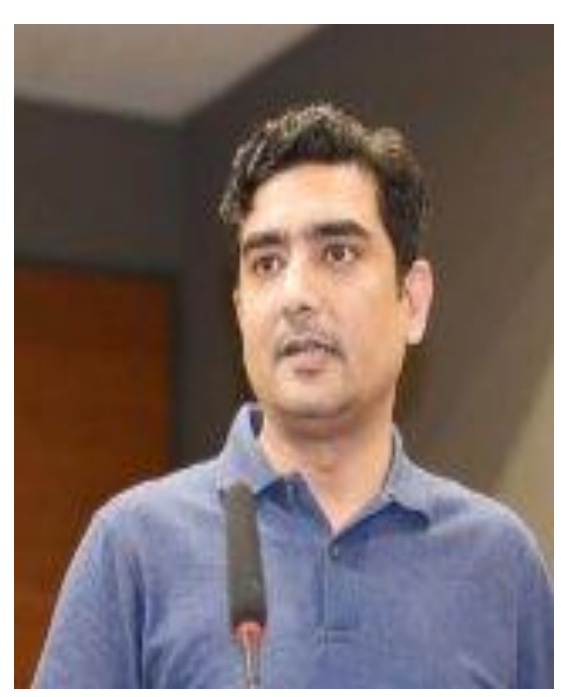

**Bacha Rehman** received the Ph.D. degree in computer science from Universiti Brunei Darussalam, Brunei, in 2019.

He is currently a Senior Lecturer in computing (AI and data science) with Solent University, Southampton, U.K. He previously served as Assistant Professor and Chairperson at Namal University, Pakistan, and held academic positions at GIK Institute, Pakistan. His research interests include machine learning, computer vision, human-robot interaction, and machine perception. He was awarded the Best Paper Award at the 2nd ICISPC (2018), received the UBD FIC-19 research grant, and earned recognition at the FG-2020 ABAW Competition.